# Security Testbed for the Internet of Things

Shachar Siboni
CSRC, Ben-Gurion University
of the Negev (BGU)
Beer-Sheva, 84105, Israel
sibonish@post.bgu.ac.il

Vinay Sachidananda
iTrust, Singapore University of
Technology and Design (SUTD)
Singapore 487372
sachidananda@sutd.edu.sg

Asaf Shabtai
CSRC, Ben-Gurion University
of the Negev (BGU)
Beer-Sheva, 84105, Israel
shabtaia@post.bgu.ac.il

Yuval Elovici
iTrust, Singapore University of
Technology and Design (SUTD)
Singapore 487372
yuval_elovici@sutd.edu.sg

## ABSTRACT
The Internet of Things (IoT) is a global ecosystem of information and communication technologies aimed at connecting any type of object (thing), at any time and in any place, to each other and to the Internet. One of the major problems associated with the IoT is maintaining security; the heterogeneous nature of such deployments poses a challenge to many aspects of security, including security testing and analysis. In addition, there is no existing mechanism that performs security testing for IoT devices in different contexts. In this paper, we propose an innovative security testbed framework targeted at IoT devices. The security testbed supports both standard and context-based security testing, with a set of security tests conducted under the different environmental conditions in which IoT devices operate. The requirements and architectural design of the proposed testbed are discussed, and the testbed operation is demonstrated in several testing scenarios.

## CCS Concepts
• **Security and privacy**➔Systems Security➔Vulnerability management • Computing methodologies➔Machine learning.

## Keywords
Internet of Things (IoT), Security, Privacy, Testbed Framework.

## 1. INTRODUCTION
The Internet of Things (IoT) consists of the combination of physical objects with sensors, actuators, and controllers with connectivity to the public world via the Internet. The low cost of hardware, along with the prevalence of mobile devices and widespread Internet access, has made the IoT a part of modern everyday life. An exponential increase in the use of the IoT is expected in the future; as it does, security issues must increasingly be considered given that all IoT devices are connected to the Internet, providing the means for hackers to obtain access to these devices.

SHODAN [1], the IoT search engine, shows the dark side of connected IoT devices, where several vulnerabilities have been discovered using this tool [2] [3]. Different Internet connected devices, ranging from cameras to industrial controllers, can be easily manipulated [4] [5]. These studies confirm both: 1) the fact that IoT devices, by their very nature, are prone to attacks, and 2) the need to seriously consider security measures for such devices. Furthermore, no common security standard exists for all IoT devices. Although there is a need to address the security challenges of the IoT ecosystem, a flexible method for evaluating the security of IoT devices does not currently exist, and there is a lack of dedicated testbeds to perform security testing and analysis on IoT devices [6].

The development of a testbed to perform comprehensive security testing and analysis for IoT devices under real conditions will help to remedy this situation. We propose a fully functional IoT testbed for security analysis in which various IoT devices such as smart home devices and smart wearables, as well as wireless sensor networks (WSNs) are tested against a set of security requirements. The IoT testbed emulates different types of testing environments which simulates various sensors activity (GPS, movement, Wi-Fi, etc.) and performs predefined and customized security tests. The testbed also collects data while performing the security test, used to conduct security forensic analysis. Furthermore, the report produced contains the type of IoT device, connectivity, communication protocols supported, and security test cases executed and their status (PASS or FAIL).

The testbed consists of hardware and software components for experiments of wide-scale testing deployments. A variety of test cases are provided by the proposed testbed such as standard, contextual, data, and side channel tests. Thus, the security testbed must support a range of security tests, each aimed at different aspect of security. Standard security testing will be performed based on vulnerability scans and penetration test methodology, in order to assess and verify the security level of IoT devices under test. (See Appendix A for a list of supported tests.) In addition, advanced security testing will be performed by the security testbed using different arrays of simulators.

Given the fact that the vast majority of security technologies adopted today are primarily focused on alerting users about specific technical aspects of an attack, rather than the root cause of an attack, the implementation of automated detection processes can be difficult. Moreover, defining the requirements for the development and implementation of such a testbed will also be a challenging task. Therefore, in this paper we divide these requirements into functional and non-functional requirements. The testbed architecture and design presented in this paper is a layer-based platform model with a modular structure. Based on this any type of IoT device can be tested in the proposed security testbed framework, including smart appliances, smart city devices, smart wearable devices, and more. In addition, any relevant simulator and/or measurement and analysis tool can be deployed in the testbed environment in order to perform comprehensive testing in the testbed.

The main contributions of this paper are threefold:
- We propose a set of requirements that are required for building an IoT security testbed.
- We propose a novel testbed architecture which is modular and adaptable.
- We present various test cases evaluated in our testbed.

The structure of the paper is as follows: after providing the introduction to the testbed in Section 1, related work is discussed in Section 2. In Section 3, we propose the necessary requirements for building the testbed. In Section 4, we introduce the testbed architecture and design which includes the security modules used for testing purposes. Section 5 provides several examples of security tests conducted using our proposed testbed, and we conclude in Section 6.

## 2. RELATED WORK

Several testbeds have been proposed for IoT devices [6]. In addition, there are a few labs around the world that focus on IoT security [7]. Most of the recent work on IoT testbeds tends to focus on a single technology domain (e.g., WSNs) [8] [9] [10] [11]. Others take a more heterogeneous approach to the study of IoT testbeds [12] [13]. There are very few studies using various IoT devices and focusing on multiple technology domains [14].

MoteLab [8], provides a testbed system for WSNs, was one of the first testbeds developed. Still in use today, it has also served as the basis for various other testbeds such as INDRIYA [15]. Kansei [9] is one of the most surveyed testbeds, which provides advanced various functions, including co-simulation support, mobility support using mobile robots, event injection possible mote level. CitySense [10] is a public mesh testbed deployed on light poles and buildings. Two features make this testbed particularly interesting: 1) its realism and domain specificity provided by a permanent outdoor installation in an urban environment, and 2) the realization of the control and management plane based solely on wireless links. The Senselab [11] testbed consists of more than 1000 sensor nodes with energy measurement supported for every node and repeatable mobility via electric toy trains. In [12] the testbed consists of federation architecture, co-simulation support, topology virtualization, in-situ power measurements on some nodes, and mobility support. FIT IoT-LAB [13] provides a very large scale infrastructure facility suitable for testing small wireless sensor devices and heterogeneous communicating objects. The testbed offers web-based reservation and tooling for application development, along with direct command-line access to the platform. However, all of the abovementioned IoT testbeds focus solely on WSNs.

The T-City Friedrichshafen [14] testbed considers various IoT devices, making it multi-domain; operated by Deutsche Telekom, it combines innovative information and communication technologies, together with a smart energy grid, to test out innovative healthcare, energy, and mobility services. Although the T-City Friedrichshafen testbed is multi-domain, it fails to take into account security aspects.

INFINITE [16], the Industrial Internet Consortium approved testbed, encompasses all of the major technologies, domains, and platforms for industrial IoT environments, covering the cloud, networks, mobile, sensors, and analytics. Projects such as FIESTA-IoT [17] provide a blueprint experimental infrastructure for heterogeneous IoT technologies. The FIESTA-IoT project consists of various testbeds like SmartCampus [18] and SmartSantander [19]. SmartSantander proposes a unique city-scale experimental research facility for common smart city applications and services. In [20], authors propose ASSET (Adaptive Security for Smart Internet of Things in eHealth), a project to develop risk-based adaptive security methods and mechanisms for IoT in eHealth. The project proposes a testbed to accurately evaluate adaptive security solutions in realistic simulation and use case scenarios, however the project does not address multi-domain IoT devices and security aspects.

Stanford's Secure Internet of Things Project [7] is a cross-disciplinary research effort between computer science and electrical engineering faculty at Stanford University; the University of California, Berkeley; and the University of Michigan. The research effort focuses on three key questions: analytics, security, and hardware and software systems. Though the project is focused on securing IoT devices, the testbed to secure IoT devices has not yet been discussed in [7].

Hence, based on our knowledge, critical gaps exist, and a testbed that focuses on security testing for IoT devices, and especially considering different context environments, has not yet been developed.

## 3. SYSTEM REQUIREMENTS

The requirements for a security testbed for the Internet of Things can be classified and formulated on various abstraction levels. The highest abstraction level reflects the security objectives. An example of a security objective could be "The system must be reliable and scalable for various scenarios."

Security requirements can stem from end-users' needs, prioritized risk scenarios, regulation laws, and best practices and standards. The system requirements section provides an overview of functional and non-functional requirements. The functional requirements include the behavioral requirements for a system to be operational, while the non-functional requirements describe the key performance indicators.

### 3.1 Functional Requirements

The functional requirements are the conditions or capabilities needed in the system to ensure the fulfilment of the testbed. For example, the tests supported, test definition, analysis of the test results, etc. Moreover, the functional requirements pertain to expected system inputs and the outputs to be produced, and the relationships between those inputs and outputs. Furthermore, these requirements describe the series of steps that are needed in order for the testbed to be operational, ranging from initializing the test to producing the test reports. Table 1 presents a concise list of functional requirements for a security testbed.

**Table 1. Functional Requirements of the Security Testbed**

| Functional Requirements | Description |
|---|---|
| Action initialization | Ability to simulate real world conditions and initialize the testing process. |
| Detection of IoT devices | Ability to detect all of the IoT devices within the testbed. |
| Adding/removing a test case | Ability to add/remove a test case in the testbed (test cases are different types of security vulnerabilities). |
| Automatically running a test case. | Ability to run the test case automatically with minimal or no intervention for all connected devices. |
| Logging the status of each test case | Ability to log the status of each test case in real-time. |
| Report generation | Ability to generate a report for all test cases executed in the testbed. |

### 3.1.1 Initialization and Detection

One of the primary functional requirements of the IoT testbed is to establish a realistic environment for the various tests performed. The IoT devices within the testbed should perform their usual tasks, as they are intended to do. By using the simulators, stimulators, and any other tools needed, the testbed should simulate real world conditions in order to test the IoT devices in different contexts.

After initialization and activation of the IoT device, the next requirement is the detection of the IoT device present in the testbed environment. During the detection process, a log file should be created consisting of the IoT device OS, the processes running, actions being performed, etc. This information will be used for any subsequent anomaly detection. The detection process should also be used in case scenarios involving compromised IoT devices that are present within the testbed environment in order to perform attacks on other IoT devices.

### 3.1.2 Security Tests

The IoT testbed must supports a range of security tests, each targeting a different security aspect. The testbed should detect various vulnerabilities that IoT devices can be prone to and provide analysis and proof of concept to these vulnerabilities. Accordingly, a security testbed should takes into account some of the vulnerabilities from OWASP [21], including:

- **Injection:** The IoT devices are prone to injection flaws such as SQL, OS, and LDAP during command or query.
- **Broken Authentication and Session Management:** IoT devices are easily compromised with implementation flaws such as passwords, session tokens, etc. [2].
- **Cross-Site Scripting (XSS):** When untrusted data is received by an IoT device and sent to a browser, XSS flaws can occur.
- **Security Misconfiguration:** Secure configurations must be defined for the IoT devices and secure settings should be implemented and maintained, particularly in cases such as smart homes which may be connected to a variety of IoT devices.
- **Sensitive Data Exposure:** While communicating with IoT devices or when two or more IoT devices communicate with each other, the data needs to be protected and the communication layer requires extra protection such as encryption at rest or in transit.
- **Missing Function Level Access Control:** The IoT device needs to perform the access control checks on the server, to prevent attackers from forging requests in order to access functionality without proper authorization.
- **Using Components with Known Vulnerabilities:** Data loss or server takeover can be facilitated by an attack when vulnerable components in the IoT device are exploited; this, in turn, can undermine IoT defenses and enable attacks.

In addition, the testbed should support templates of tests and scenarios. The testbed should be capable of running automated tests based on specific requirements (e.g., extract all tests that are relevant to the accelerometer sensor) or the device type (e.g., all tests that are relevant to IP cameras). Moreover, the tests within the testbed should be executed in a sequential manner and with an appropriate structure. Furthermore, the testbed should provide a success criteria for each test (for example, binary pass/fail or a scale from 1-[pass] to 5 [fail], which may be based on a predefined threshold provided by the system operator in advance). Moreover, the success criteria should not be generic and should be defined for a specific tested IoT device and/or tested scenario. The system must be able to evaluate the test results against the test criteria.

### 3.1.3 Logging and Analysis

After conducting a series of functional requirement steps, the testbed should be capable of logging the tests. The system collects various data during the test execution, including network traffic information (e.g., about Wi-Fi, Bluetooth, and ZigBee operation), IoT device internal status information (e.g. CPU utilization, memory consumption, and file system activity), etc. This information should be stored as a log file for further analysis.

In addition, the testbed system should support intelligent analysis. For some tests the system operator should be able to define a decision rule specifying whether the device passed the test or not. For example, if the existing vulnerability test module identifies high risk vulnerability the device fails the test, etc. Such rules should be defined for specific types of IoT devices, data sources, and testbed capabilities. In some cases it is impossible to define a hard decision rule and the testbed should expose the results of analysis modules (e.g. anomaly detection model) through a user interface for manual exploration and decision.

## 3.2 Non-Functional Requirements

The non-functional requirements are the set of attributes which characterize the IoT testbed. Therefore, these requirements tend to be related and can be derived from functional requirements. The non-function requirements are as follows.

### 3.2.1 Usability

Usability ensures the testbed's ease of use, without the need for extensive efforts on the part of the user. Security testbed should be easy to operate and use (with easily defined tests, easy to input configuration and to interpret output).

### 3.2.2 Security-Related

- **Reliability:** is the ability of the IoT testbed to perform its required functions under the stated conditions for a specific period of time. Furthermore, reliability can be considered as: (a) the availability of the IoT testbed service when requested by end-users, and (b) the failure rate which is how often the IoT testbed fails to deliver the service requested by the users.
- **Anti-Forensic:** is the capability of the testbed to detect and subsequently prevent malicious applications on the IoT device (if it has been infected) from being activated. Malicious applications, in particular, tend to disturb the forensic analysis operation, and the testbed should be able to prevent the IoT device from being activated to overwrite any sensitive data that shouldn't fall into the wrong hands.
- **Security:** is the ability of the testbed to ensure authorized access to the system. To safeguard the integrity of the IoT testbed from accidental or malicious damage, security requirements should maintain: 1) access permissions, where the testbed data may only be changed by the system administrator; 2) backup of all testbed data in database; and 3) encrypted communication between the different components and parties of the testbed.

- **Accountability (including non-repudiation):** is the capability of the testbed to keep audit records to support independent review of access to resources/uses of capabilities – not only the data collected, but also the log files must be specified and protected.
- **Controlled:** is the ability of the testbed to prevent malicious IoT devices from being activated.

### 3.2.3 Adaptive

The security testbed should be able to adapt in accordance with new application domain concepts and support various communication types.

- **Scalability:** is the capability of the testbed to increase total throughput under an increased load when resources (typically hardware) are added.
- **Performance:** is the ability of the speed of operation of the testbed. Performance requirements pertain to: 1) throughput requirements which define how much the testbed can accomplish within a specified amount of time, and 2) response requirements which define how quickly the testbed reacts to user input.
- **Flexibility:** is the capability to modify the testbed after deployment. This includes: 1) adaptability, the ability of the testbed to be adapted based on new requirements and application domain concepts; 2) sustainability, the ability to fix faults and deal with new technology; and 3) customizability, the capability of the testbed configuration to be customized and fine-tuned by the user.

## 4. TESTBED ARCHITECTURE AND DESIGN

In this section, the architecture and design of the proposed security testbed for IoT devices are presented. This includes an in-depth description of the testbed's modules and system components (both software and hardware based).

### 4.1 System Architecture

The functional architecture model of the security testbed, illustrated in Figure 1, is designed based on the requirements described in Section 3. It is a layer-based platform model with a modular structure, which means that any type of IoT device can be tested in the proposed security testbed framework, including smart appliances, smart city devices, smart wearable devices, and more. In addition, any relevant simulator and/or measurement or analysis tool can be deployed in the testbed environment. A detailed description of the modules that comprise the functional model and the interactions between these modules as a complete security testing system are provided. Note that the architecture model suggested here is based on our existing model, as the current work is a continuation of research in this subject.

### 4.1.1 Management and Reports Module (MRM)

This module is responsible for a set of management and control actions, including starting/initializing the test, enrolling new devices, simulators, tests, measurement and analysis tools, and communication channels, and generating the final reports upon completion of the test. The testbed operator (the user) interfaces with the testbed through this module using one of the communication interfaces (CLI\SSH\SNMP\WEB-UI) in order to initiate the test, as well as to receive the final reports. Accordingly, the module interacts with the Security Testing Manager Module and the Measurements and Analysis Module, respectively. The MRM contains a system database component that stores all relevant information about the tested device (including the OS, connectivity, sensor capabilities, advanced features, etc.), as well as stores information regarding the test itself (including config files, system snapshots, and test results).

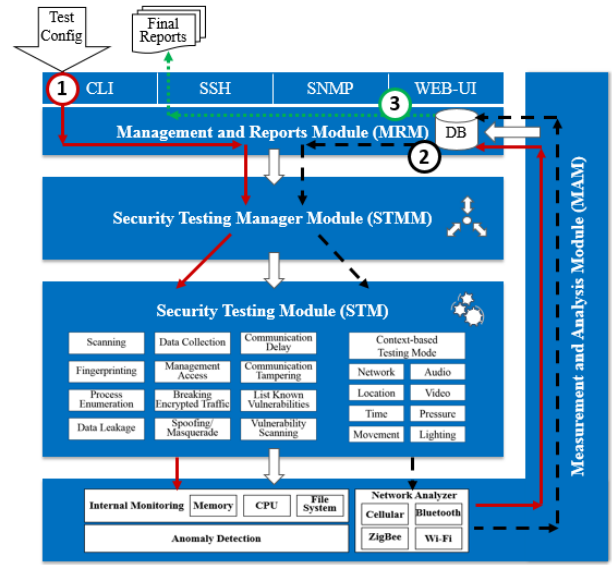

**Figure 1. Security testbed framework - abstract functional architecture model.**

### 4.1.2 Security Testing Manager Module (STMM)

This module is responsible for the actual testing sequence executed by the security testbed (possibly according to regulatory specifications). Accordingly, it interacts with the Security Testing Module in order to execute the required set of tests, in the right order and mode, based on predefined configurations provided by the user (based on the config file loaded in the MRM).

### 4.1.3 Security Testing Module (STM)

This module performs standard security testing based on vulnerability assessment and penetration test methodology, in order to assess the security level of the IoT device under test (DUT). See Appendix A for a list of supported tests and the appropriate test/success criteria for each test. The STM is an operational module which executes a set of security tests as plugins, each of which performs a specific task in the testing process. The module also supports a context-based testing mode, where it generates various environmental stimuli for each sensor/device under test. Meaning, in this mode of operation, the STM simulates different environmental triggers and runs the security tests, in order to identify and detect context-based attacks that may be launched by the IoT-DUT. This is obtained using a simulator array list, such as a GPS simulator or Wi-Fi localization simulator (for location-aware and geolocation-based attacks), time simulator (using simulated cellular network, GPS simulator, or local NTP server), movement simulator (e.g., using robots), etc. See Table 2 for a list of supported simulators. The module interacts with the Measurements and Analysis Module in order to monitor and analyze the test performed.

### 4.1.4 Measurements and Analysis Module (MAM)

This module employs a variety of measurement (i.e., data collection) components and analysis components (both software and hardware-based). The measurement components include different network sniffers for communication monitoring such as Wi-Fi, cellular, Bluetooth, and ZigBee sniffers, and device monitoring tools for measuring the internal status of the devices under test. The analysis components process the collected data and evaluate the results according to a predefined success criterion. Note that most of the predefined success criteria are not generic and are defined for a specific tested IoT device and/or tested scenario. In some cases, a success criterion cannot be clearly defined, therefore advanced analysis tools and mechanisms will be deployed in the testbed (for example, a network-based anomaly detection tool will be employed to process the recorded network traffic of the tested IoT device in order to detect anomalous events in the system). In this case, the pass/fail decision will be based on a predefined threshold provided by the system operator in advance. The detected anomalies should then be investigated and interpreted by the system operator using dedicated exploration tools which are part of the user interface.

**Table 2. Simulators Supported by the Security Testbed**

| Simulator | Description |
| --- | --- |
| Network | The testbed uses network simulators to simulate different network environments, such as Wi-Fi, Bluetooth, ZigBee, and more, in order to support different network connectivity in the testbed. |
| Location | The testbed simulates different locations and trajectories using the GPS generator device, in order to test the behavior of the IoT device under test in different locations/trajectories. |
| Time | The testbed simulates different days of the week and times of day using either the GPS generator device, internal NTP server, or internal cellular network, in order to test the behavior of the IoT device under test at different times. |
| Movement | The testbed simulates different movements using either robots or human testers, in order to test the behavior of the IoT device under test while performing different movements. |
| Lighting | The testbed simulates different lighting levels, in order to test the behavior of the IoT device under test in different lighting scenarios. |
| Audio | The testbed simulates audio using a voice simulator, in order to test the behavior of the IoT device under test in different sound environments. |
| Video | The testbed simulates images, pictures, and videos using a video simulator, in order to test the behavior of the IoT device under test during different video changes. |

### 4.1.5 Testing Process

The testing process shown in Figure 1 starts by loading a configuration file (by the user/testbed operator) in the testbed via the MRM component. Based on the configuration loaded, a set of security testing is conducted in the testbed (illustrated as Phase 1 on the red line in Figure 1) using the STM component. The results are then stored in the system database component. Next, context-based security testing is performed using the STM component (illustrated as Phase 2 on the black dashed line in Figure 1), by selecting the appropriate simulators for the test. In this phase, different simulators are employed in order to realistically simulate the environment in which IoT devices operate, and the same set of security tests are conducted (again, based on the configuration file loaded in advance). The results obtained are then stored in the system database component. Both of these testing phases are controlled by the STMM component. Note that during the execution of the testing process, different measurement and analysis tools are employed using the MAM component, in order to collect relevant information about the test performed (including network traffic, internal status of the IoT-DUT, etc.). Finally, a forensic analysis is performed by the MRM component, based on the results obtained from both phases and the information collected during the testing process. The final results of the overall testing process are then generated and sent to the user/testbed operator (illustrated as Phase 3 on the green dashed line in Figure 1).

## 4.2 System Components

The testbed environment, illustrated in Figure 2, includes both software and hardware system components. From the internal software system component perspective, this includes the user interface and several testbed managers' modules, each responsible for a specific task. From the environmental system component point of view, this includes the IoT device under test (IoT-DUT), the set of security test tools, measurement, and analysis tools, and a set of simulator/stimulator devices employed in the testbed, as described next.

### 4.2.1 Internal Software System Components

The internal software system components of the security testbed include the user interface (GUI/Remote), testbed manager (orchestrator), test manager, element manger, and storage manager components, which will be discussed next.

#### 4.2.1.1 User Interface – GUI/Remote

The user interface component is used for sending/receiving commands/test results to/from the testbed, respectively. This can be handled locally (e.g., using a GUI) or remotely (e.g. REST API). SSH and Telnet connectivity are supported as well.

#### 4.2.1.2 Testbed Manager – Orchestrator

The testbed manager component acts as an orchestrator in the system. It is responsible for managing the workflow between the software system components of the testbed (including the underlying managers: Element Manager, Test Manager, etc.). This component also exposes the interface of the testbed, to the user as well as to the hardware system components of the testbed.

#### 4.2.1.3 Test Manager

The test manger component is responsible for the creation and execution of testing scenarios. A scenario is an object of the system that reflects the testing process. Therefore, it manages test objects, each composed of a set of action objects (for more details see Section 4.3). Accordingly, the test manager creates and manages scenario, test, and action objects; generates templates for each of these objects, and is responsible for executing the testing scenario, as well.

#### 4.2.1.4 Element Manager

The element manager component is responsible for provisioning and deleting elements from the testbed. An element is a general term used in the testbed that applies to both software and hardware. Each element is defined by its driver. A driver is a programmable component that exposes the element's capabilities, either to the user or to other elements of the testbed. Types of elements in the testbed are: IoT-DUT, simulators/stimulators, measurement and analysis tools, and security tests.

#### 4.2.1.5 Storage Manager

The storage manager component is a repository of system elements. In addition, it is responsible for writing and logging different events occurring in the system, before, during, and after the test is conducted (e.g., registering simulator, driver event, test action being running, test results, etc.).

### 4.2.2 Environmental System Components

The environmental system components include both hardware and software components, including: the IoT device under test, a set of security tools, environmental simulators and stimulators,

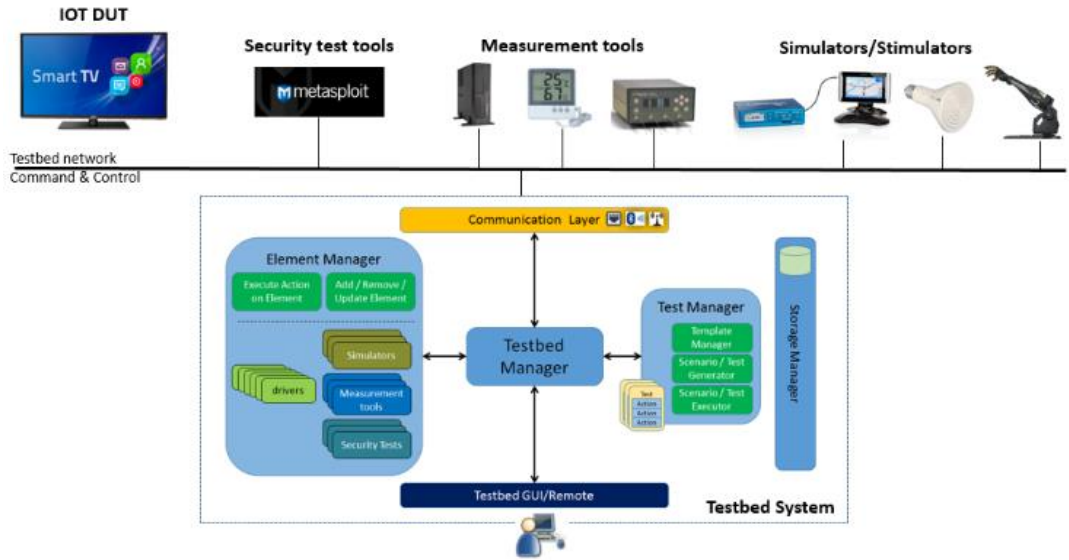

**Figure 2. IoT Security Testbed System Components.**

and different types of measurement and analysis tools, as discussed next.

*4.2.2.1 IoT Device Under Test (IoT-DUT)*

The security testbed is designed and implemented to support examination of a wide range of IoT devices, including different categories such as: smart home appliances, smart industrial equipment, smart city devices, wearable devices, and more.

*4.2.2.2 Security Test Tools*

The security testbed utilizes different security testing tools available online, including the Nmap security scanner tool for network discovery and security auditing [22], the Wireshark tool for network protocols analysis [23], Aircrack [24] to assess WiFi network, Metasploit is used for penetration testing [25], all these tools running under the Kali Linux penetration testing environment [26]. Nessus [27], OpenVAS [28], Cain-and-Abel [29], and OSSEC [30] security tools will be employed in the testbed as well.

*4.2.2.3 Measurement and Analysis Tools*

The security testbed uses different types of measurement and analysis tools, including: data collection modules, analysis and security rating modules, data analysis modules, and more. These modules will be developed in order to meet the needs of the testbed; for example, different anomaly detection models will be used in the testbed in order to automatically identify and detect anomalies both in the network traffic and in the internal status (CPU, memory, file system, system calls, etc.) of the IoT-DUTs.

*4.2.2.4 Simulators and Stimulators*

The security testbed employs different types of environmental simulators and stimulators (e.g., a GPS simulator that simulates different locations and trajectories, movement simulators such as robotic hands, etc.). Using the provided set of simulators (simulator array), the testbed realistically generates arbitrary real-time stimulations, ideally for all sensors of the tested IoT devices. See Table 2 for a list of the simulators supported by the testbed.

### 4.3 Testing Scenarios Methodology

The creation of a testing scenario in the testbed is based on the object definitions shown in Table 3. First, a *scenario* object should be created using the test manager component of the testbed. A *scenario* object represents a specific testing scenario (test case) in the system, and it is defined as a complete and final set of *test* objects. Once a *scenario* object is defined, it may be saved as a scenario template for future use. A *test* object, which defines a specific test in the system, is a complete and final set of *action* objects. Once a *test* object is defined, it may be saved as a test template for future use. An *action* object represents a specific testing operation that needs to be performed. The *action* object includes an *initiator*, which is the element in the system that is responsible for initiating the action (this is done manually by the user or automatically by one of the system's components, such as a simulator, measurement tool, etc.), the *element* on which the action is done/referred to, a *command* to be executed in this action, and the required input *parameters* (params) for this specific command/action, as described in Table 3. Note that an action is executed by one of the elements of the system, on a specific element of the system. In addition, each action object generates a trace log of that action in the system. All of these objects (scenarios, tests and actions), are the building blocks for the creation of a testing scenario in the testbed and are managed by the test manager component of the testbed. Using this testing methodology, the tests within the testbed are structured and executed in a sequential manner.

### 4.4 Possible Extensions for the Testbed

The testbed is designed and implemented as a plugin framework in order to support future operational capabilities. For example, one of the possible extensions for the testbed is an IoT honeypot plugin/module which will employed as a security tester element in the system. This module will interface with an IoT honeypot environment in order to collect data about attacks learned and patterns observed within the IoT honeypot framework. The module will maintain a database of attacks for specific IoTs and will generate these attacks in the testbed. Using this extension, the testbed can be used as a physical sandbox for different IoT devices. Another possible extension is a risk assessment and management plugin/module which will employed as a security analysis element in the system. This module will be used to calculate the probability of attacks and their severity of impact in order to quantify risks for different IoT-DUTs. The incorporation of these plugin extensions (and others), will enhance the testbed's capabilities, and ensure that the testbed will serve as a comprehensive security testing platform now and in the future, for future IoT case scenarios.

**Table 3. Object definitions used to create testing scenarios**

| Object | Description |
|---|---|
| Action | Syntax: <Initiator, Element, Command, Params><br>• Initiator: The initiator of an action (either handled manually by the user or automatically by one of the system's entities).<br>• Element: The element (component) on which the action is done/referred to, including: testing environment element (e.g., simulators/stimulators, measurement and analysis tools, etc.), the user (or testbed operator), and the IoT device under test.<br>• Command: The command to be done in this action, including: START, STOP, CREATE, DELETE, MODIFY, SET, TEST, NOTIFY, SELECT, REMOVE, LOGIN, TEST_CONNECTION, etc.<br>• Params: The required input parameters (params) for this specific command/action.<br><br>Example: <USER, GPS_SIM, START, {trajectory.cfg}>. |
| Test | Represents a specific test in the system, defined as a complete and final set of action objects. |
| Scenario | Represents a specific testing scenario in the system, defined as a complete and final set of test objects. |

## 5. TESTBED IMPLEMENTATION AND OPERATION

In this section we describe the testbed implementation and present several examples of its operation. We deployed the testbed system in an isolated room inside our lab (shown in Figure 3), which provides a testing environment with minimal external disruptions. We integrated the NI TestStand testing environment [31], and this software system provided the testbed's testing management infrastructure. TestStand served as an orchestrating tool in the testbed and was used to create testing scenarios based on the testing methodology discussed in subsection 4.3. After a scenario is defined, TestStand can execute and manage the test sequence, as well as control and operate all selected components of the testbed, including security tools (e.g., Nmap), simulators (e.g., GPS simulator), and measurement and analysis tools (e.g., Wireshark) for the given scenario. TestStand is also used to send requests to the user (for user intervention purposes), and to generate the final report upon test completion. TestStand also serves as an interface with the NI LabVIEW tool [32] in order to integrate different hardware testing components and advanced capabilities into the testbed, as described below. We now present several test scenarios for the testbed's operation.

### 5.1 Test Scenario 1: Context-based Testing

In this test case we demonstrate the testbed's operation in a context-based scenario. For that matter, we implemented a case scenario in which a compromised smartwatch device executes a network mapping attack on an organizational network once the location and/or the time of the day the attacker set in advance are identified (note that as part of the attack, the compromised smartwatch device is directly connected to the organizational Wi-Fi network). Accordingly, we established the appropriate testing environment in the testbed for this test scenario and executed the test. A Wi-Fi network simulator (Cisco router device) was used to simulate the organizational network environment, and a GPS simulator (LabSat 3 device) was used to simulate the changes in locations and time of the day in order to trigger the attack. Furthermore, Wireshark (running on a sniffer device in the testbed) and a device monitoring tool that we implemented (running via a tester device in the testbed) were employed during the testing process in order to monitor both the network traffic and the internal status of the smartwatch device under test (the CPU, memory, and file system), respectively. The setup configuration for this test scenario is shown in Figure 4.

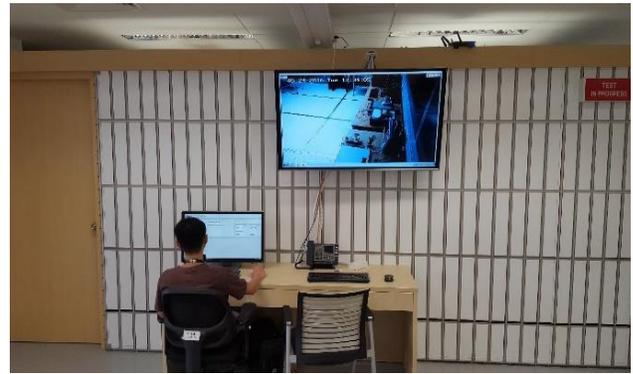

**Figure 3. The isolated room in which the testbed was deployed, equipped with an internal IP camera in order to control the testing process outside the testbed environment.**

The testing sequence executed within the TestStand testing environment is shown in Figure 5. First, all of the relevant components of the testbed for the test were automatically configured via TestStand; these include the IoT-DUT, network sniffer, Wi-Fi and GPS simulators, and the data collection tools. The test was then executed by TestStand. In this phase of the testing process a prerecorded path was replayed in the testbed environment by the GPS simulator, and both the sniffer and the tester devices recorded the network traffic, the CPU and memory status of the smartwatch device under test, and the GPS signals, respectively. Finally, upon completion of the test, we gathered all of the test results, including the pcap file, the internal status of the DUT, and more, and manually analyzed the results obtained.

Based on the analysis conducted (illustrated in Figure 6), we were able to detect two network mapping attacks, as well as one false alarm event which occurred during the testing process. Moreover, we managed to identify the exact locations and the time of the day in which the attacks executed. We accomplished this by correlating the anomalous events between the tested smartwatch's activities (CPU and memory), the network traffic behavior (observed in the testbed environment), and the GPS measurements (time and location) which recorded during the test.

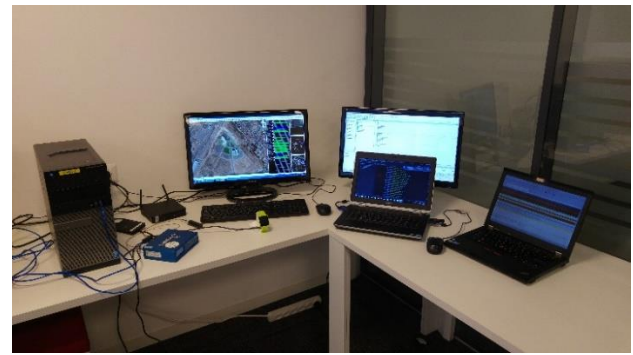

**Figure 4. The testbed configuration used for the context-based scenario, including GPS and Wi-Fi simulators, Tester and Sniffer monitoring devices, the smartwatch device as the IoT-DUT, all controlled and managed by the NI TestStand tool.**

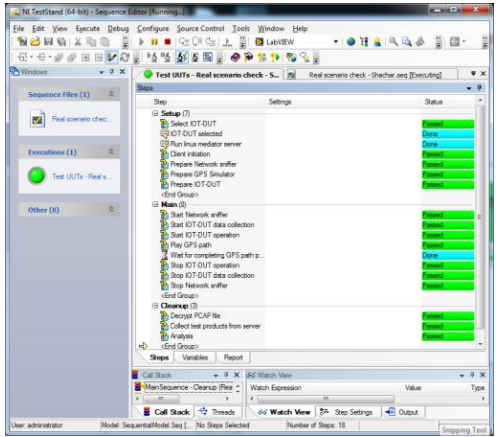

**Figure 5. The testing sequence executed in NI TestStand tool for the context-based scenario.**

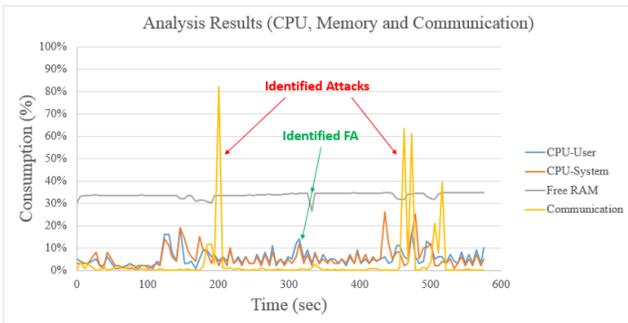

**Figure 6. The results for the context-based testing scenario, including the internal status of the smartwatch device under test (CPU and memory perspectives), and the communication traffic observed in the testbed environment during the test.**

## 5.2 Test Scenario 2: Profiling IoT Devices

In this test case we demonstrate the testbed's operation for profiling IoT devices in the testbed environment. For that matter, we implemented a profiler module and integrated it into the testbed system as a tester/analyzer tool. The profiler module is based on a machine learning algorithm which analyzes the network traffic observed during the testing process. The module is operated in two phases, a training phase and a testing phase. During the training phase of the test scenario, a statistical model is trained based on the network traffic captured in the testbed environment. Given this statistical model, different IoT devices are analyzed in the testbed (based on their network activity as well) during the testing phase, in order to automatically profile the tested devices. For more details about the profiler module implementation and operation see Appendix B.

This case scenario shows the testbed's ability to automatically identify and profile IoT devices based only on their network activity. Using this software component we can learn how IoT devices behave. For example, we can identify whether a specific malicious IoT device impersonates other IoT devices, based on observing the network traffic during its operation. Moreover, we can use this component in order to categorize IoT devices (both benign and malicious) into behavioral groups, defining groups of devices based on specific behavioral measures.

The testing process conducted for this test scenario is as follows. For the training phase, first we deployed several IoT devices in the testbed, including a smart thermostat device, a movement sensor, smartwatch and smart glass (Google Glass) devices, IP camera, a smart socket, and a smart TV, as shown in Figure 7. Next, we created a testing scenario in TestStand, in order to record the network activity of these devices in the testbed environment during this phase of the test. This information was then used by the profiler module to define the statistical model in the testbed. For the testing phase of this test scenario, we deployed other IoT devices in the testbed environment, including a smart home kit (a Piper device: the All-in-One wireless security system for the home), a smartwatch device (a different device than that used in the training phase), and a smart connected lighting kit (a Philips Hue kit, for connecting ZigBee bulbs in the testbed, that was analyzed based on the communication interaction between the application installed in a smartphone device with the smart bulb bridge device of that kit), as shown in Figure 7. We created another testing scenario in TestStand, where in this phase of the test, these IoT devices were tested against the constructed statistical model and profiled respectively (meaning, these IoT devices were profiled using the profiler module based on their network activity observed during this phase of the test). See Figure 8 for the testing sequence executed in the TestStand testing environment for this test scenario.

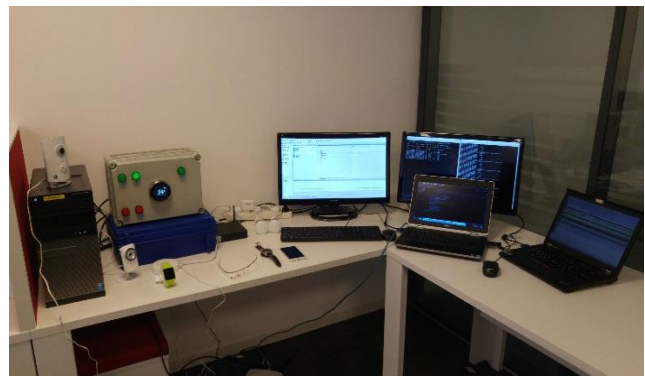

**Figure 7. The testbed configuration used for the profiling IoT devices scenario, including the array of simulators (Wi-Fi router and IoT devices), several IoT-DUTs, and a sniffer monitoring device, all controlled and managed by the NI TestStand tool. The profiler module integrated as a software component in TestStand for this test scenario.**

Note that the IoT devices that were used during the training phase were employed as environmental simulators in the testbed, in order to establish a more complex (yet realistic) environmental IoT setting; the statistical model was constructed based on these devices. Whereas, the IoT devices that were used in the testing phase are the IoT-DUTs for this specific test scenario (these devices were profiled based on the statistical model constructed in the training phase). Moreover, to be able to monitor the network activity of all of the IoT devices, each of them was connected to the Wi-Fi network that was simulated in the testbed, and a sniffer device was used (during both the training and the testing phases).

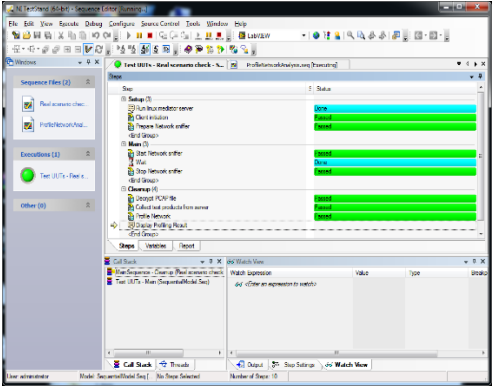

**Figure 8. The testing sequence executed in NI TestStand tool for the profiling IoT devices test scenario.**

Upon completion of the test, the final results for this test scenario (the decision made by the profiler module for each tested IoT device given the statistical model), were generated in the TestStand environment. Table 4 presents the results obtained. Note that the rows in the table refer to the IoT devices that were used to construct the statistical model during the training phase, and hence are defined as the testing model in the table; the columns in the table refer to the IoT devices that were tested during the testing phase of this test scenario, and hence are defined as IoT-DUTs in the table. The numbers in the table are the estimations of the statistical model for each IoT-DUT given the model (with respect to each IoT device that was used to construct the statistical model). As can be seen from the table, the tested smartwatch device is profiled as a Smartwatch device, given the statistical model, with a level of accuracy of 85.71%. This means that by using the profiler module we able to automatically identify and profile the smartwatch device under test in the testbed with a relatively high level of accuracy. The smart bulb bridge device, however, was profiled as a Smart TV (with 56.53% accuracy) and a Smartwatch device (with 21.68%), given the model; and the smart home kit device (Piper) was profiled as a Smart Glass device (with 72.90% accuracy).

**Table 4. Profiling IoT-DUTs given a statistical model**

| Testing Model | IoT-DUTs | | |
| --- | --- | --- | --- |
| | Smart Home Kit | Smart Bulb Bridge | Smartwatch |
| Thermostat | 0.00% | 0.03% | 0.00% |
| Movement Sensor | 0.00% | 0.10% | 0.00% |
| Smartwatch | 20.00% | 21.68% | 85.71% |
| Smart Glass | 72.90% | 0.78% | 13.43% |
| IP Camera | 3.81% | 1.02% | 0.00% |
| Smart Socket | 0.00% | 5.81% | 0.00% |
| Smart TV | 0.48% | 56.53% | 0.00% |

The results can be explained as follows. As long as the statistical model includes instances of the same types as the tested devices, the profiler module will automatically identify and profile the tested devices with a high degree of accuracy (as in the case of the smartwatch device under test). This means that once the testbed (using the profiler module) learned about the activity of an IoT device (its network activity), it is possible to successfully (or effectively) profile other IoT devices from the same family (type). However, if the statistical model does not already include instances of the same type as the tested devices (as in the cases of the smart home kit and smart bulb bridge devices), the tested devices will be profiled by the profiler module as devices that most closely match instances that are included in the statistical model. Moreover, for the final calculations we used a weighted average with respect to the probability estimation obtained from the statistical model (where predictions were estimated with some level of percentage of accuracy, not necessary 100%), therefore the overall profiling of each tested IoT device (the sum of the column of each IoT-DUT) is not 1. Note that the main objective of this case scenario is to illustrate the testbed's ability to automatically profile IoT devices. In future work, we intend to enhance the profiler module in order to improve its level of accuracy. This includes training more IoT devices in the testbed, and in different context-based scenarios, in order to collect additional information about the real network traffic of IoT devices, and to employ other machine learning algorithms in order to establish a more accurate statistical model.

### 5.3 Test Scenario 3: Port Scanning

In this test case, we show the penetration testing methodology, port scanning and show the complete flow of our testbed. The motivation behind using port scanning as a test case is to demonstrate how an IoT device in an environment can be vulnerable with open ports and how an IoT device can be affected.

The setup for this test case is that we have a PC that runs the LabVIEW, to make sure that we can connect various IoT devices (as IoT-DUT's), measurement tools, etc., as shown in Figure 9. In this scenario the Power Supply Unit (PSU) is connected to the LabVIEW PC and the IoT-DUT is the IP camera. The setup also contains another PC which runs as the analysis machine and testing tool Nmap is present.

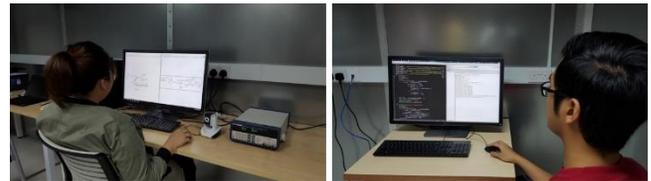

**Figure 9. The setup showing the PC running LabVIEW and connected to the PSU and the IP camera as IoT-DUT (top left picture). PC running Nmap (top right picture).**

In this test scenario, the sequence of steps (starting from initializing the test to generating the report), is orchestrated from TestStand. TestStand works like a flow control and coded to fit required task to test IoT devices. The TestStand initializes the test and sends a message to LabVIEW to activate. PSU is connected to LabVIEW which is in turn connected to the IP camera as IoT-DUT. LabVIEW controls the power supply and is able to turn ON and OFF the IoT-DUT. Accordingly, in the TestStand test sequence, the voltage and the ampere input are requested from the user (via two pop ups, one asking a value for voltage and another for ampere) and the entered values are stored as the local variables. TestStand sends the entered values as the variants to LabVIEW, where the variants are decoded and the values are appended to the respective location to set voltage and ampere. LabVIEW will then turn ON the PSU, which turns ON the IP camera connected to the PSU.

Once the IoT-DUT is ON, it automatically connects to the dedicated experimental access point. Additional wait time has been included to allow the IoT-DUT to connect to the access point. Set of Python scripts is developed in the analysis machine for further interaction with the TestStand. After the wait time expires, the Pyton script in the analysis machine triggers the TestStand for the IP address of the IoT-DUT to be tested. After entering the IP Address to scan, TestStand will trigger analysis machine to run the Nmap to discover the open ports on the entered IP Address through a SSH setup. After Nmap finishing the port scan, the results are saved as an XML file. TestStand will then trigger the analysis machine to run a custom Python script that extracts the discovered open ports from the XML file. The XML file is looped line by line checking for the keyword "Discovered". Any line detecting the keyword "Discovered" is appended in a file with list of open ports. Finally, we compare our scan results with the score list of top vulenrable ports [33]. The file with the score list of top vulenerable ports contain the top vulnerable ports number, descriptions of the ports and the metric score given to each port. The Python script compares the open port against a list of vulnerable open ports and identify those vulnerable ports for reporting. If the word "Discovered" is not found in the XML file, the whole XML file is copied as the output result, which indeed displays everything that is scanned.

We have setup a metric system to evaluate the risk level of open ports. The risk level is set as: 0 – safe, <15 – minor risk, 15< && <30 – major risk and >30 – critical risk. The Python script in the analysis machine saves two text files to return to TestStand. The "Overall Results" and the "Metric Score". As shown in Figure 10, after obtaining the scan results from the Nmap, the scan results are compared with the score list of top vulnerable ports to provide the "Overall Results". In the "Overall Results" the list of open ports, ports that are considered to be vulnerable and the metric ratings are reflected. For e.g., in one of our test case the list of open ports discovered were 135/tcp, 139/tcp, 80/tcp, 5900/tcp, 445/tcp, 443/tcp, 49152/tcp, 6646/tcp, 2869/tcp on IP 10.0.8.100. The ports that were considered vulnerable with services running were 80: A web server is running on this port with Score: 3, 5900: A vnc server is running on this port with Score: 3, 445: Microsoft-DS Active Directory, Windows shares with Score: 1, 443: A TLSv1 server answered on this port with Score: 5 and 49152: The Win32 process 'wininit.exe' is listening on this port with Score: 1. In this case, the overall Risk Level is assessed as Minor Risk and the Metric Score is 13.

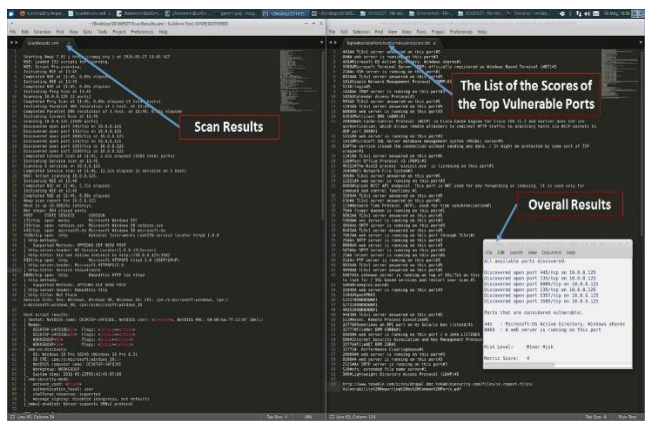

**Figure 10. The results obtained from the port scanning test scenario.**

At this point in process, TestStand will then trigger LabVIEW to determine the "Risk Factor" of the IoT Device. It will call on the "MetricScore" file, retrieve the metric number, and determine the "RISK" from a pre-defined "Risk Margin". After LabVIEW obtains the "Risk Factor" string, the information is sent back to TestStand. Finally, TestStand will then move on to the step of turning OFF the PSU. After turning OFF the PSU, TestStand will then show the conclusion of the test through a popup message. Log file is automatically saved in a fixed location for user access.

# 6. DISCUSSION AND FUTURE WORK

The Internet of Things (IoT) is an emerging technology that transforms ordinary physical devices, such as televisions, refrigerators, watches, cars, and more, into smart connected devices. The potential applications associated with the IoT are seemingly infinite, with new and innovative features and capabilities being developed almost daily. However, the extensive benefits and opportunities provided by IoT computing are accompanied by major potential compromises in data privacy and security, in that any smart device becomes a security risk. Moreover, due to the heterogeneous nature of such devices (different OSs installed, produced by different manufacturers, several communication channels supported, etc.) and the fact that they are used in a variety of contexts, analyzing and ensuring the security of such devices is considered a complex task.

Therefore, in this paper we propose an innovative security testbed framework for IoT devices. The proposed testbed is designed to perform traditional security testing, based on penetration test methodology, in different contexts and environments. This is accomplished by realistically simulating the environment in which IoT devices exist (such as location, movement, etc.), in order to identify and detect context-based attacks that may be carried out by compromised IoT devices.

The proposed security testbed aims for a black-box approach in which we assume that only the final product is available. In addition, the proposed framework is targeted specifically at IoT devices and is designed to execute relevant security tests with minimal human intervention. The downside of such an approach is that it cannot provide a mapping of the IoT device (or its functions) at a specified security/assurance level, but instead provides a list of the test results (based on the success criteria defined for each test in Appendix A). Moreover, detecting context-based attacks requires the execution of a security test in various contexts. We can assume that simulating all possible contexts in the testbed is not feasible due to the potentially large number of context variables. For example, when considering location as a context for the security test, although we can use a simulation application to generate different trajectories and replay them in the testbed (e.g., using SATGEN GPS simulation software [34]), it will be impossible to run a context-based test that covers all possible locations. Therefore, we define two types of context-based tests: *targeted* and *sample* tests. In a *targeted* test we assume that a bounded set of contexts to be evaluated by the testbed is provided as an input to the testing process. For example, an IoT device that is going to be deployed in a specific organizational environment will be tested with the organization's specific geographical location, given the execution limits of the testbed. In a *sample* test, a subset of all possible contexts (those that can be simulated) is evaluated. This subset is selected randomly according to a priori assumptions about contexts of interest (e.g., malicious activity is usually executed at night, the device is installed in a home environment, etc.).

In future work we intend to enhance the testbed's system capacity in order to support its full operational capability. This includes deployment of additional simulator devices, implementation of advanced measurement and analysis tools, and further automation of the testing process. Moreover, in order to extend the scope of the security testing we intend to connect the security testbed with a honeypot environment. Based on that, additional requirements for a satisfactory IoT security testbed, as well as its potential limitations, will be provided. This will allow us to define which features are essential for testing various IoT devices used in different contexts and in different environments.

# APPENDIX

# A. PENETRATION TESTS SUPPORTED BY THE SECURITY TESTBED

The following table presents a list of penetration tests and an appropriate test/success criterion for each test that are supported by our proposed security testbed.

**Table 5. Penetration Tests Supported by the Security Testbed**

| Test | Description | Test/Success criteria |
|---|---|---|
| Scanning (e.g., IP and port scanning) | Investigate the detectability of IoT devices by observing wireless/wired communication channels. Attempt to identify the existence of the device. Enumerate communication channels/traffic types observed, open ports, etc. | *Undetectable*- the IoT-DUT cannot be detected by the testbed via any communication channel; *Safe*- the IoT-DUT is detectable, but no open ports were observed; *Minor risk*- the IoT-DUT is detectable, and common ports are open, e.g., port 80 (HTTP), 443 (HTTPS), etc.; *Major risk*- the IoT-DUT is detectable, and uncommon ports for such devices are open, e.g., ports 20, 21 (FTP), port 22 (SSH), port 23 (Telnet), etc.; or, *Critical risk*- the IoT-DUT is detectable, and unexpected ports are open in the device. |
| Fingerprinting | By monitoring communication traffic to/from the device, attempt to identify the type of device, its operating system, software version, list of all sensors supported, etc. | *Unidentifiable*- the type of IoT-DUT cannot be identified by the testbed; *Safe*- the device provides identifiable information, but all the IoT-DUT's software versions are up-to-date; *Minor risk*- some low risk detected applications, e.g., calendar, etc., are out-of-date; *Major risk*- some major risk detected applications, e.g., navigator, mail, etc., are out-of-date; or, *Critical risk*- operating system and critical applications are out-of-date. |
| Process enumeration | Lists all running processes on the device and presents their CPU and memory consumptions. This can be done by monitoring the device's activities, e.g., using ADB (Android Debug Bridge) connectivity. | *Safe*- the list of processes cannot be extracted without admin privileges; *Moderate risk*- the list of processes can be extracted without admin privileges on the device only; or, *Fail*- the list of processes can be remotely extracted without admin privileges. |
| Data leakage | Validate which parts of the communication to/from the device are encrypted (and how) or sent in clear text, and accordingly check if an application leaks data out of the device. | *Pass*- traffic is encrypted, and no data leaks are detected; or, *Fail*- traffic is unencrypted and sent in clear text, therefore data may leak from the IoT-DUT. |
| Data collection | Check if an application on an IoT device collects sensor data and stores it on the device. This can be achieved by monitoring the locally stored data and correlating sensor events. | *Safe*- the tested application does not collect and store data on the IoT-DUT; *Minor risk*- the tested application collects and stores normal data, e.g., multimedia files, on the IoT-DUT; *Major risk*- the tested application collects and stores sensitive data, e.g., GPS locations, on the IoT-DUT; or, *Critical risk*- the tested application collects and stores critical information, e.g., device status (CPU, memory, sensor events, etc.), on the IoT-DUT. |
| Management access | Attempt to access the management interface/API of a device using one of the communication channels. Access could be obtained by using default credentials, a dictionary attack, or other known exploits. | *Pass*- management access ports, e.g., port 22 (SSH), port 23 (Telnet), are closed; or, *Fail*- one of the management access ports is open on the tested device. |
| Breaking encrypted traffic | Apply known/available techniques of breaking encrypted traffic. For example, try to redirect HTTPS to HTTP traffic (SSLstrip) or impersonate remote servers with self-certificates (to apply a man-in-the-middle attack). | *Pass*- unable to decrypt traffic sent/received by/to the IoT-DUT with the applied techniques; or, *Fail*- able to decrypt traffic data sent/received by/to the IoT-DUT using the applied techniques. |
| Spoofing/masquerade attack | Attempt to generate communication on behalf of the tested IoT device. For example, determine if any of the communication types can be replayed to the external server. | *Pass*- reply attack failed; or, *Fail*- replay attack successful. |
| Communication delay attacks | Delay the delivery of traffic between the device and remote server, without changing its data content. Determine which maximal delays are tolerated on both ends. | *Safe*- the time delay between two consecutive transactions of the IoT-DUT is within the *defined/normal* range; or, *Unsafe*- the time delay is greater than the *defined/normal* range. |
| Communication tampering | Attempt to selectively manipulate or block data sent to/from the device. For example, inject bit errors on different communication layers or apply varying levels of noise on the wireless channel. | *Safe*- the device ignores received manipulated/erroneous data; or, *Unsafe*- the device crashes or behaves unexpectedly when manipulated/erroneous data is sent. |
| List known vulnerabilities | Given the type, brand, version of the device, running services, and installed applications—list all known vulnerabilities that could be exploited. | *Safe*- no relevant vulnerabilities were found; *Minor risk* - insignificant/low risk vulnerabilities were found; or, *Unsafe*- significant and critical vulnerabilities where found. |
| Vulnerability scan | Search for additional classes of vulnerabilities by: (1) utilizing existing tools (or developing new dedicated ones as part of the ongoing research) that attempt to detect undocumented vulnerabilities such as buffer overflow and SQL injection; (2) maintaining a database of attacks (exploits) detected on previously tested IoTs or detected by honeypots, and evaluate relevant/selected attacks on the tested IoT; and (3) using automated tools for code scanning. | *Safe*- no new vulnerabilities were found during the testing process conducted; *Minor risk*- insignificant/low risk new vulnerabilities were found; or, *Unsafe*- significant and critical new vulnerabilities were found. |

## B. PROFILER MODULE FOR IOT DEVICES

The profiler module for IoT devices is based on a machine learning algorithm which analyzes the network traffic observed in the testbed environment during the testing process. As mentioned in subsection 5.2, the module operates in two phases, a training phase and a testing phase, as illustrated in Figure 11. First, in the training phase, a statistical model is defined based on the network activity recorded during this phase (for that matter, several IoT devices are deployed in the testbed environment). Next, in the testing phase, different IoT devices (that were not used during the training phase) are profiled based on their network behavior, given the statistical model constructed in previous phase.

Note that the training phase is executed initially only once, although it may be necessary to retrain the statistical model later (in order to update the model with new IoT devices). The testing phase is executed each time we test (profile) an IoT device in the testbed. Moreover, the IoT devices used during the training phase were employed as environmental simulators in the testbed, whereas the IoT devices that were used in the testing phase were operated as the IoT-DUTs for this case scenario. Furthermore, in both phases, a sniffer device records the network traffic observed in the testbed environment during the test (which represented as pcap files and used as an input for the testing process).

used to construct the statistical model during the training phase (see Figure 12 for the result confusion matrix of the statistical model), and the latter is used to profile each IoT device under test during the testing phase, given the statistical model constructed in the training phase. Both submodules based on the features-based discrete sequences generated in each phase. All of the above submodules were wrapped in a c# code in order to implement the profiler module as a complete tester/analyzer component in the testbed, as well as to enable integration of the profiler module in the NI TestStand testing environment.

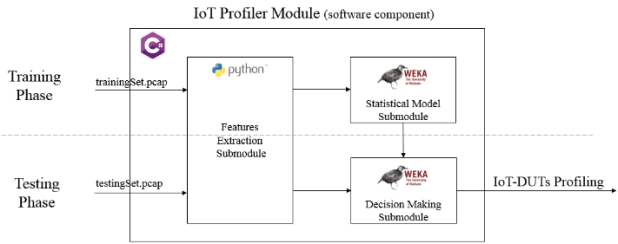

**Figure 11. The IoT profiler module structure, including features extraction, statistical model and decision making submodules.**

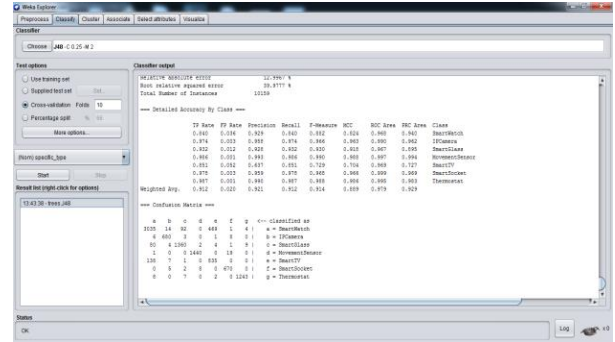

**Figure 12. The result confusion matrix of the constructed statistical model.**

The profiler module is composed of several submodules, as presented in Figure 11. A features extraction submodule implemented in python is used to represent the recorded network traffic (during both the training and the testing phases) as discrete sequences of network traces, based on a set of selected features (e.g., session-based features, such as ttl, pkt_size, pkt_arrival_time, etc.). The statistical model submodule and the decision making submodule are both based on the implementation of the J48 classification algorithm in Weka [35]. The former is